# Differential Phase-contrast Interior Tomography

Wenxiang Cong and Ge Wang

Biomedical Imaging Division, School of Biomedical Engineering and Sciences, Virginia Tech, Blacksburg, VA 24061, USA

wcong@vt.edu; wangg@vt.edu

**Abstract:** Differential phase contrast interior tomography allows for reconstruction of a refractive index distribution over a region of interest (ROI) for visualization and analysis of internal structures inside a large biological specimen. In this imaging mode, x-ray beams target the ROI with a narrow beam aperture, offering more imaging flexibility at less ionizing radiation. Inspired by recently developed compressive sensing theory, in numerical analysis framework, we prove that exact interior reconstruction can be achieved on an ROI via the total variation minimization from truncated differential projection data through the ROI, assuming a piecewise constant distribution of the refractive index in the ROI. Then, we develop an iterative algorithm for the interior reconstruction and perform numerical simulation experiments to demonstrate the feasibility of our proposed approach.

## 1. Introduction

The current x-ray computed tomography (CT) technology relies on the attenuation contrast mechanism. However, biological soft tissues in preclinical and clinical imaging mainly consist of atoms of low atomic numbers, such as hydrogen, carbon, nitrogen and oxygen, and their elemental composition is nearly uniform. The attenuation differences among different light elements in biological tissues are quite small, leading to an indiscernible contrast in attenuation-based x-ray CT images. By contrast, x-ray phase-contrast imaging promises to improve image contrast significantly for a wide range of biological and medical studies [1-3]. It utilizes the refractive mechanism underlying the interaction between x-rays and biological tissues. The refractive index can be expressed in a complex form: $n(\mathbf{r}) = 1 - \delta(\mathbf{r}) + i\beta(\mathbf{r})$, where $\delta(\mathbf{r})$ denotes the refractive index decrement at a voxel, which is related to a phase-shift, $\beta(\mathbf{r})$ characterizes the attenuation strengh of the material. The interaction cross section of the x-ray phase shift is about one thousand times larger than that of the attenuation for an x-ray energy range of 10–100 keV, which implies that phase-contrast imaging has a much higher sensitivity for light elements compared to conventional attenuation-contrast imaging. Therefore, x-ray phase-contrast imaging can be used to observe structural variations of biological tissues, such as vessels, lesions, and fibrous tissues, and enhance the sensitivity and specificity relative to traditional x-ray imaging while potentially reducing radiation damage to the patient or animal [4-5].

Unlike attenuation-based conventional local tomography, interior tomography aims at the theoretically exact solution to the interior problem, and has been actively studied over the past several years. In this imaging mode, x-rays irradiate a region of interest (ROI) only, representing a long-standing challenge. The general interior problem does not have a unique solution only from truncated projection data [6]. Interestingly, recent results show that the interior problem is solvable if appropriate yet practical prior information is available. In particular, if the attenuation coefficient distribution on a small subregion in an ROI is known [7-9], or the attenuation coefficient distribution on the ROI has some characteristic, such as being piecewise constant or polynomial [10-12], the interior problem does have a unique solution.

Differential phase-contrast interior tomography we propose here is defined as the theory and methods to reconstruct a refractive index distribution over an ROI in a theoretically exact fashion. The previous results in this area are relevant but have not touched the goal of what we defined for differential phase-contrast interior tomography. Anastasio et al. presented differential phase-contrast tomographic imaging of an ROI from truncated differential projection data using a backprojection filtration (BPF) algorithm [13]. Pfeiffer et al. reported region-of-interest computed tomography using filtered backprojection (FBP) method based on grating-based x-ray differential phase-contrast imaging [14]. In practice, gratings of large sizes are difficult to fabricate and model, and acquired projection data may only cover a small portion of an object. While the recent interior tomography papers reported encouraging results, the current theory and techniques cannot be directly applied for differential phase-contrast interior tomography, because the involved phase-contrast data are truncated differential projection data, instead of projection data themselves.

In this paper, we demonstrate that the differential phase-contrast interior problem has a unique solution if an underlying x-ray refractive index distribution is piecewise polynomial in an

ROI, and the piecewise constant refractive index distribution over the ROI can be reconstructed via the total variation minimization from truncated differential projection data through the ROI.

## 2. Differential Phase-contrast Interior Tomography Theory

When a parallel x-ray beam passes through an object, the phase shift data can be expressed as a projection of the refractive index distribution in the object [15]:

$$R[\delta(s,\theta)] = \frac{2\pi}{\lambda} \int_R \delta(s\theta + t\theta^\perp) dt, \tag{1}$$

where $\lambda$ is the x-ray wavelength, $\theta$ a directional vector of the x-ray beam at a projection angle $\alpha$, and $\theta = (\cos\alpha, \sin\alpha)$. In x-ray grating interferometry imaging or diffraction enhanced imaging (DEI), one can obtain differential projection data $\partial_s R[u(s,\theta)]$ by measuring the Moiré fringe pattern. Without loss of generality, in this paper we consider the 2D case of differential phase-contrast interior tomography. An object is supported on a disk $\Omega_A = \{\mathbf{r} = (x,y) \in R^2 : |\mathbf{r}| < A\}$, and an interior ROI defined as $\Omega_a = \{\mathbf{r} = (x,y) \in R^2 : |\mathbf{r}| < a\}$ for $0 < a < A$. Using a similar method to attenuation interior tomography [12], we can obtain following Lemma 1,

**Lemma 1**: If $h(\mathbf{r})$ and $\delta(\mathbf{r})$ are piecewise smooth for $\mathbf{r} \in R^2$, compactly supported on the disk $\Omega_A$, and have same differential phase projection data $\partial_s R[h(s,\theta)] = \partial_s R[\delta(s,\theta)], -a \leq s \leq a, \theta \in S^1$, then the difference function $u(\mathbf{r}) = h(\mathbf{r}) - \delta(\mathbf{r})$ is analytic in the disk $\Omega_a$, and $\partial_s R[u(s,\theta)] = 0, -a \leq s \leq a, \theta \in S^1$.

**Lemma 2**: If a function $u(x)$ is (1) a polynomial function for $x \in (-a,a)$; and (2) $Hu(x) = 0$ for $x \in (-a,a)$, where $Hu(x)$ is the Hilbert transform of $u(x)$, then $u(x) = 0$ almost everywhere.

**Proof**: Let $\Omega = (C \setminus R) \cup (-a, a)$, $g(x) = Hu(x)$ for $x \in (-\infty, \infty)$ and define

$$f(x) = -\frac{1}{\pi} \left( \int_{-\infty}^{-a} \frac{g(t)}{z-t} dt + \int_{a}^{\infty} \frac{g(t)}{z-t} dt \right) \text{ for } z \in \Omega. \tag{2}$$

The function $f(z)$ is analytic in $\Omega$. Let $z = x + iy$ with $y > 0$, and we have following formulas from Eq. (2):

$$\text{Re}[f(z)] = \frac{1}{\pi} \int_R \frac{t-x}{(t-x)^2 + y^2} g(t) dt, \text{ and } \text{Im}[f(z)] = \frac{1}{\pi} \int_R \frac{-y}{(t-x)^2 + y^2} g(t) dt \tag{3}$$

From $f(z) - f(\bar{z}) = 2i\text{Im}[f(z)]$, we obtain [16-17]

$$\lim_{y\to 0^+}[f(z)-f(\bar{z})]=\frac{1}{2i}g(x) \quad \text{almost everywhere.} \tag{4}$$

From the above Condition (1) and (2), we have that $f(z)$ is a polynomial function in $\Omega$. Hence, we have $g(x)=0$ almost everywhere from Eq. (4), which implies $u(x)=0$ almost everywhere.

**Lemma 3.** If an image $u(\mathbf{r})$ satisfies: (1) $u(\mathbf{r})=p(\mathbf{r})$ for $x\in\Omega_a$, where $p(\mathbf{r})$ is a 2D polynomial function, and (2) $\partial_s R[u(s,\theta)]=0, -a\leq s\leq a, \theta\in S^1$, then $u(\mathbf{r})=0$.

**Proof**: For an arbitrary $\varphi_0\in[0,\pi)$, let $L_{\theta_0}$ be the line through the origin and tilted by $\theta_0=(\cos\varphi_0,\sin\varphi_0)$. When $u(\mathbf{r})$ is restricted to the line $L_{\theta_0}$, it can be expressed as

$$v_{\varphi_0}(t)=u(t\cos\varphi_0,t\sin\varphi_0),\quad t\in(-\infty,\infty). \tag{5}$$

By the relationship between the backprojection of differentiated projection data and the Hilbert transform of an image [12,18], we have

$$Hv_{\varphi_0}(t)=-\frac{1}{2\pi}\int_{\varphi_0-\frac{\pi}{2}}^{\varphi_0+\frac{\pi}{2}}\partial_s R[u(s,\theta)]\big|_{s=\mathbf{r}\cdot\theta}d\varphi, \tag{6}$$

where $\mathbf{r}=(t\cos\varphi_0,t\sin\varphi_0)$, and $\theta=(\cos\varphi,\sin\varphi)$. By Condition (1) and Eq. (5), we have $v_{\varphi_0}(t)=p(t\cos\varphi_0,t\sin\varphi_0)$, $t\in(-a,a)$, where $p(t\cos\varphi_0,t\sin\varphi_0)$ is a polynomial function with respect to t. By Condition (2) and Eq. (6), we have

$$Hv_{\varphi_0}(t)=0 \quad \text{for } t\in(-a,a). \tag{7}$$

By Lemma 2, we have that $v_{\varphi_0}(t)=0$, which implies that $u(\mathbf{r})=0$ almost everywhere.

From Lemma 3, we obtain the following result:

**Theorem 1.** Suppose that a refractive index image $\delta(\mathbf{r})$ of an object is piecewise polynomial in $\Omega_a$. If another refractive index image $h(\mathbf{r})$ is also piecewise polynomial in $\Omega_a$ and has the same differential phase projection data as that of $\delta(\mathbf{r})$, $\partial_s R[h(s,\theta)]=\partial_s R[\delta(s,\theta)]$ for $s\in(-a,a), \theta\in S^1$, then $h(\mathbf{r})=\delta(\mathbf{r})$.

Threorem 1 show that the interior reconstruction of refractive index from truncated differential projection data has a unique solution within the class of piecewise polynomial functions in $\Omega_a$. Next, we present a practical method to implement the interior reconstruction to find a accurate refractive index image.

In practical numerical computation, we commonly discretize the region of iterest $\Omega_a$ into numerical mesh, for example, dividing $\Omega_a$ into finite rectangle sub-regions:

$G = \{E_k, \ k = 1, 2, \cdots, N\}$ such that $\Omega_a \subset \bigcup_{k=1}^{N} E_k$. Based on the numerical discretization of $\Omega_a$, the total variation (TV) of a function $h(x, y)$ is defined by

$$\mathrm{TV}(h) = \sum_{k=1}^{N} D_{E_k}(h) \tag{8}$$

$D_{E_k}(h)$ represents the anisotropic norm:

$$D_{E_k}(h) = |h(x_k, y_k) - h(x_k + d, y_k)| + |h(x_k, y_k) - h(x_k, y_k + d)|, \tag{9}$$

or isotropic norm:

$$D_{E_k}(h) = \sqrt{[h(x_k, y_k) - h(x_k + d, y_k)]^2 + [h(x_k, y_k) - h(x_k, y_k + d)]^2}, \tag{10}$$

where $(x_k, y_k)$, $(x_k + d, y_k)$, and $(x_k, y_k + d)$ are nodal coordinates of $E_k$ with mesh size of $d$. From the definition of TV, we have following results.

**Theorem 2.** For a sufficient small size of numerical mesh, the total variation (TV) of a piecewise constant function in $\Omega_a$ is smaller than that of any other functions, if they have the same truncated differential phase projection data through the ROI.

**Proof**: We assume that $\Omega_a$ contains a finite number of subsets, $\Omega_a = \bigcup_{m=1}^{M} \Omega_m$, with $\delta(\mathbf{r}) = c_m$, for $\mathbf{r} \in \Omega_m, m = 1, 2, \cdots, M$. We denote the boundary between sub-regions $\Omega_i$ and $\Omega_j$ as $\Gamma_{ij} = \overline{\Omega}_i \cap \overline{\Omega}_j$. We seperate the set $G$ into two parts, $G = B \cup C$, $B$ only includes rectangles which interact with the boundary $\Gamma_{ij}$ in region $\Omega_a$ and $C$ consists of rectangles which locates inside subsets $\Omega_m, m = 1, 2, \cdots, M$. Let $\delta(\mathbf{r})$ is a piecewise constant function in $\Omega_a$ and its truncated differential phase projection data of ROI is $\partial_s R[\delta(s, \theta)]$ for $s \in (-a, a)$, $\theta \in S^1$. For any other function $h(\mathbf{r})$, which has the same truncated differential phase projection data to that of $\delta(\mathbf{r})$, we have,

$$\sum_{E_k \in G} D_{E_k}(h) = \sum_{E_k \in B} D_{E_k}(h) + \sum_{E_k \in C} D_{E_k}(h). \tag{11}$$

Let $u(\mathbf{r}) = h(\mathbf{r}) - \delta(\mathbf{r})$. Because $\delta(\mathbf{r})$ is piecewise constant in $\Omega_a$, we obtain

$$\sum_{E_k \in C} D_{E_k}(h) = \sum_{E_k \in C} D_{E_k}(u), \tag{12}$$

and

$$\sum_{E_k \in G} D_{E_k}(\delta) = \sum_{E_k \in B} D_{E_k}(\delta). \tag{13}$$

By triangle inequality, we have

$$\sum_{E_k \in B} D_{E_k}(\delta) - \sum_{E_k \in B} D_{E_k}(u) \le \sum_{E_k \in B} D_{E_k}(h). \tag{14}$$

From Lemma 1, $u(\mathbf{r})$ is a analytic function in $\Omega_a$. Clearly, the measure of $B$ will decrease and measure of $C$ will increase for smaller size of numerical mesh. For a sufficient small size of numerical mesh, following inequality hold,

$$\sum_{E_k \in C} D_{E_k}(u) - \sum_{E_k \in B} D_{E_k}(u) \ge 0. \tag{15}$$

From Eqs. (11), (12), (14), and (15), we obtain,

$$\sum_{E_k \in B} D_{E_k}(\delta) \le \sum_{E_k \in G} D_{E_k}(h). \tag{16}$$

From Eqs. (13) and (16), we obtain the result in the theorem.

From Theorem 2, if true refractive index distribution is piecwise constant in an ROI, the differential phase-contrast interior reconstruction can be implemented through the total variation minimization in terms of the refractive index. Furthermore, in the same spirit of the above proofs, theorem 2 is easily extended to a piecewise linear (or polynomial) refractive index distribution in an ROI.

**Corollary.** For a sufficient small size of numerical mesh, the second-order total variation (TV) of piecewise linear function in $\Omega_a$ is smaller than that of any other functions, if they have the same truncated differential phase projection data through the ROI, where the second-order total variation of a function $h(\mathbf{r})$ is defined by a anisotropic TV

$$\text{TV}^2(h) = \sum_{i=1}^{M} \sum_{j=1}^{N} \left( \left| h(x_i - d, y_j) - 2h(x_i, y_j) + h(x_i + d, y_j) \right| + \left| h(x_i, y_j - d) - 2h(x_i, y_j) + h(x_i, y_j + d) \right| \right), \text{ or}$$

isotropic TV

$$\text{TV}^2(h) = \sum_{i=1}^{M} \sum_{j=1}^{N} \sqrt{\left[ h(x_i - d, y_j) - 2h(x_i, y_j) + h(x_i + d, y_j) \right]^2 + \left[ h(x_i, y_j - d) - 2h(x_i, y_j) + h(x_i, y_j + d) \right]^2},$$

here $(x_k - d, y_k), (x_k + d, y_k), (x_k, y_k - d)$, and $(x_k, y_k + d)$ are neighbor nodes of $(x_k, y_k)$ in numerical mesh with mesh size of $d$.

## 3. Numerical Simulation

Forbild phantom was employed to evaluate the proposed approach. It consisted of 40 disks. Each disk was assigned with a constant different refractive index to mimic biological tissues in the range of $[0.1 \times 10^{-5}, 0.6 \times 10^{-5}]$. A region of interest (ROI) of the phantom was assigned, as shown in Fig. 2 (a). We adoped a parallel-beam imaging geometry similar to grating interferometer imaging mode, and equi-angularly acquire 360 projections over an $180^0$ range. The detector array included 367 elements to collect the x-ray phase shift data through the ROI. The differential phase shift data was computed using the difference method from the projection data of the phantom, and corrupted by Gaussian noise to yield a truncated differential phase shift data with the signal-to-noise ratio of 10dB to mimic real measurement condition, as shown in Fig. 1. Based on Theorem 2, the interior reconstruction of refractive index can be formulated as following optimization model with total variation of anisotropic norm [19]:

$$\begin{cases} \min_{\delta} \quad \sum_{i=1}^{M}\sum_{j=1}^{N}\left(\left|\nabla_x\delta(i,j)\right|+\left|\nabla_y\delta(i,j)\right|\right) \\ s.t. \quad \left\|2\pi iw\widetilde{\mathbf{A}}_\varphi \delta - \widetilde{\mathbf{D}}_\varphi\right\|_2 \leq \varepsilon \end{cases} \quad (17)$$

where $\widetilde{\mathbf{A}}_\varphi$ is the discrete Fourier transform of the system matrix, $\widetilde{\mathbf{D}}_\varphi$ is the discrete Fourier transform of differential phase shift data $\partial_s R[\delta(s,\theta)]$, $\nabla_x \delta(i,j) = \delta(i,j) - \delta(i+1,j)$, and $\nabla_y \delta(i,j) = \delta(i,j) - \delta(i,j+1)$. The $l_1$-norm regularization is a good measure of the refractive index image sparsity, and the split Bregman iterative algorithm is a efficient method to solve the $l_1$-norm optimization [20]:

$$\begin{cases} \delta^{k+1} = \arg\min_{\delta}\left(\frac{\lambda}{2}\left\|2\pi iw\widetilde{\mathbf{A}}_\varphi \delta - \widetilde{\mathbf{D}}_\varphi\right\|^2 + \frac{\alpha}{2}\left\|\nabla_x\delta - u_x^k + v_x^k\right\|^2 + \frac{\alpha}{2}\left\|\nabla_y\delta - u_y^k + v_y^k\right\|^2\right) \\ u_x^{k+1} = \arg\min_{u}\left(\|u\|^2 + \frac{\alpha}{2}\left\|u - \nabla_x\delta^{k+1} - v_x^k\right\|^2\right) \\ u_y^{k+1} = \arg\min_{u}\left(\|u\|^2 + \frac{\alpha}{2}\left\|u - \nabla_y\delta^{k+1} - v_y^k\right\|^2\right) \\ v_x^{k+1} = v_x^k + \left(\nabla_x\delta^{k+1} - u_x^{k+1}\right) \\ v_y^{k+1} = v_y^k + \left(\nabla_y\delta^{k+1} - u_y^{k+1}\right) \end{cases} \quad (18)$$

where λ and α are regularization parameters. In the iteration, the first sub-problem only involves a least square type problem, and can be efficiently solved. The second sub-problems are the $l_1$ norm minimization, and can be solved via soft thresholding. We performed 100 iterations. As a result, the reconstructed image was in good agreement with true one inside an ROI in both image structure and pixel values, as shown in Fig. 2 (b), (d)-(e). Comparatively, we also performed a local FBP reconstruction from the same truncated differential phase shift data. We found that the structure of the reconstructed refractive index image was in good shape, but some noise, as shown in Fig. 2 (c). The major problem of local FBP reconstruction is the big shift of pixel values.

**4. Discussions and Conclusion**

In summary, we have demonstrated that the differential phase-contrast interior problem has a unique solution within the class of piecewise polynomial functions. In numerical analysis framework, we gave a rigorous proof that the accurate interior reconstruction can be achieved using TV (or high-order TV) minimization based on the truncated differential projection data. This approach allows the definition of TV with an anisotropic form and isotropic form as well, offering more flexibility in optimization algorithms. We further developed an iterative algorithm for differential phase-contrast interior tomography. Our numerical simulations have verified our theoretical finding and the feasibility of the proposed approach.

Interior x-ray Talbot interferometer imaging and interior diffraction enhanced imaging provide effiecient acquisition methods of differential phase shift data. These techniques can be widely applied for interior tomographic imaging of biological soft tissues, nondestructive testing, food inspection, archaeometry, or security sceening. The interior phase tomographic imaging allows a substantially reduction of radiation dose relative to conventional attenuation-based imaging. Furthermore, in x-ray propagation-based phase imaging, measured quantity is relatived to the second derivatives of phase shift data linked by the transport intensity equation (TIE) for an object of weak absorption and slow phase variation [21]. Using the second derivative data, interior reconstruction of a refractive index over a region of interest (ROI) is an open topic, and is still under investigation.

**Acknowledgment**: This work is partially supported by the National Institutes of Health Grant NIH/NCI CA135151 and NIH/NHLBI HL098912.

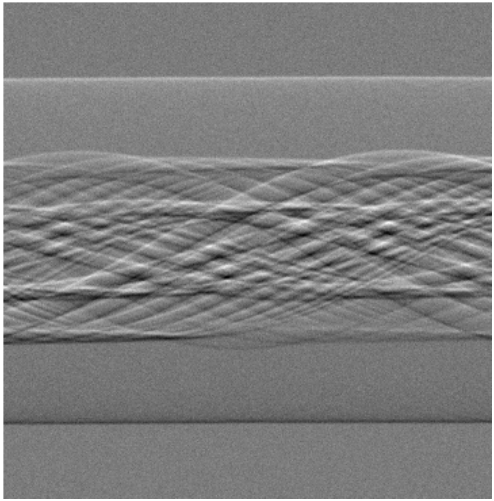

**Figure 1**. Sinogram of differential phase projection data

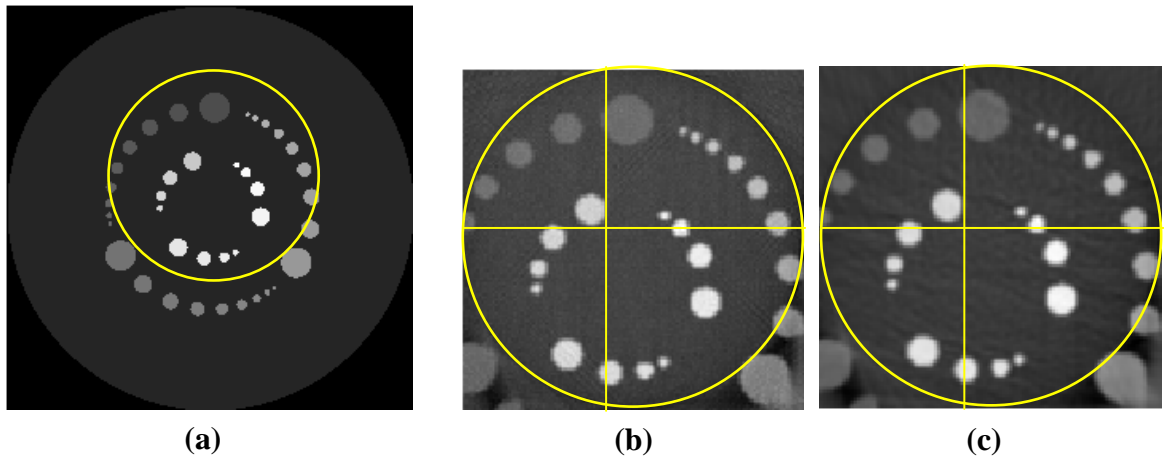

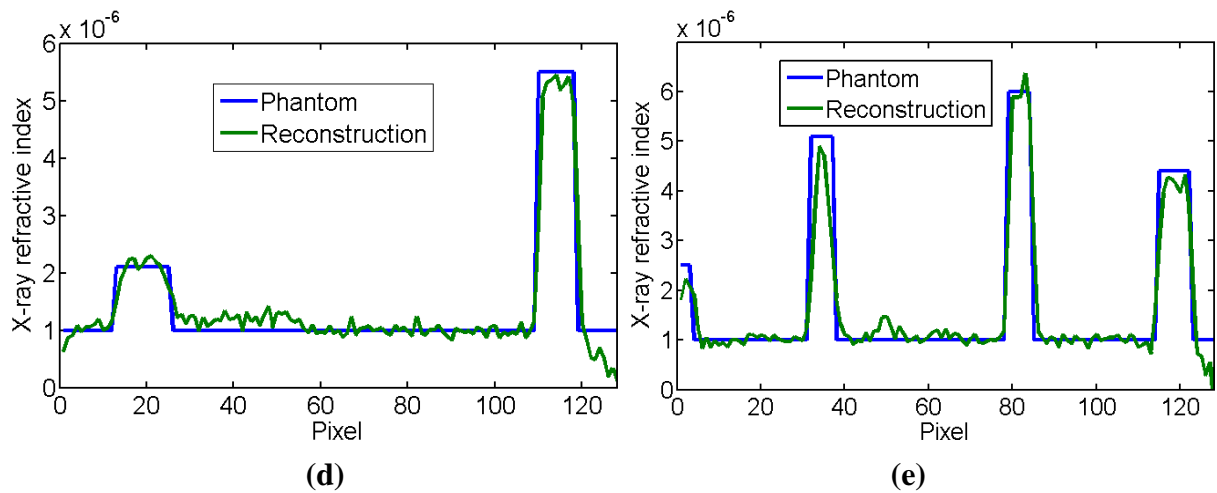

**Figure 2.** Interior reconstructed images of a numerical phantom. (a) The original phantom, (b) the reconstructed image using the proposed CS-based interior tomography algorithm, (c) the reconstructed image from local FBP method, (d) and (e) the profiles of vertical and horizontal lines corresponding to x=55 pixel and y=60 pixel, respectively.